%% file: main.tex
\documentclass[
 reprint,
superscriptaddress,
amsmath,amssymb,
]{revtex4-2}

\usepackage{float}
\usepackage{graphicx}
\usepackage{dcolumn}
\usepackage{bm}

\usepackage{makecell}

\usepackage{array}

\begin{document}

\preprint{APS/123-QED}

\title{Reconstructing social sensitivity from evolution of \\ content volume in Twitter}

\author{Sebasti\'an Pinto}
 \email{spinto@df.uba.ar}
 \affiliation{Departamento de F\'isica, FCEN, Universidad de Buenos Aires. Pabell\'on 1, Ciudad Universitaria, 1428EGA, Buenos Aires, Argentina.}
\affiliation{Instituto de F\'isica de Buenos Aires, CONICET. Ciudad Universitaria, 1428EGA, Buenos Aires, Argentina.}
\author{Marcos A Trevisan}%
 \affiliation{Departamento de F\'isica, FCEN, Universidad de Buenos Aires. Pabell\'on 1, Ciudad Universitaria, 1428EGA, Buenos Aires, Argentina.}
\affiliation{Instituto de F\'isica de Buenos Aires, CONICET. Ciudad Universitaria, 1428EGA, Buenos Aires, Argentina.}\author{Pablo Balenzuela}%
 \affiliation{Departamento de F\'isica, FCEN, Universidad de Buenos Aires. Pabell\'on 1, Ciudad Universitaria, 1428EGA, Buenos Aires, Argentina.}
\affiliation{Instituto de F\'isica de Buenos Aires, CONICET. Ciudad Universitaria, 1428EGA, Buenos Aires, Argentina.}
\date{\today}

\begin{abstract}

We set up a simple mathematical model for the dynamics of public interest in terms of media coverage and social interactions. We test the model on a series of events related to violence in the US during 2020, using the volume of tweets and retweets as a proxy of public interest, and the volume of news as a proxy of media coverage. The model succesfully fits the data and allows inferring a measure of social sensibility that correlates with human mobility data. These findings suggest the basic ingredients and mechanisms that regulate social responses capable of ignite social mobilizations.

\end{abstract}

\maketitle

\section*{Introduction}

The continuous expansion of the digital environment creates new and faster ways to exchange information and opinions \cite{Wu2007}. At the same time, it also provides access to unprecedented amounts of data, allowing the quantitative investigation of the forces that underlie the diffusion of information \cite{leskovec2009meme} and the formation of public interest \cite{Pinto2019, Albanese_2020}.

Dynamical systems have been particularly successful in identifying collective mechanisms that give rise to public opinion \cite{towers2015mass, muhlmeyer2019event}. Using variables that describe the expansions and contractions of content volume, these models explain empirical data remarkably well \cite{lorenz2019accelerating}. 

In the domain of social media, the emergence of extreme opinions that arise from moderate initial conditions has been recently disclosed \cite{baumann2020emergence, baumann2020modeling}. But extreme social reactions appear also beyond the domain of opinions and debates. Normally, people react to the news by sharing information and discussing opinions. In a few occasions however, and under heightened social sensitivity, a reactive state may emerge giving rise to street manifestations, protests and riots \cite{Drury2020} that have been extensively studied and modeled \cite{Agamennone2020,Bonnasse-Gahot2018}. 

Is it possible to extract a measure of social sensitivity from content volume coming from digital media?
Here we hypothesize that the social sensitivity regulates the dynamics of the public reacting to the media coverage of massive events. To test this hypothesis, we set up a deliberately simple model for public interest modulated by media coverage and social interactions \cite{Guo2015,castellano2009statistical, balenzuela2015undecided, barrera2020polarizing} that allows us to infer a measure of social sensitivity. We capitalize on the paradigmatic model developed by Granovetter \cite{granovetter1978threshold} based on the concept of {\em critical mass}, which represents the fraction of interested people needed to induce interest to the rest of the population. We investigate this in connection with a series of highly sensitive events that took place in the US during 2020.

\begin{figure*}[ht]
    \centering
    \includegraphics[width = \textwidth]{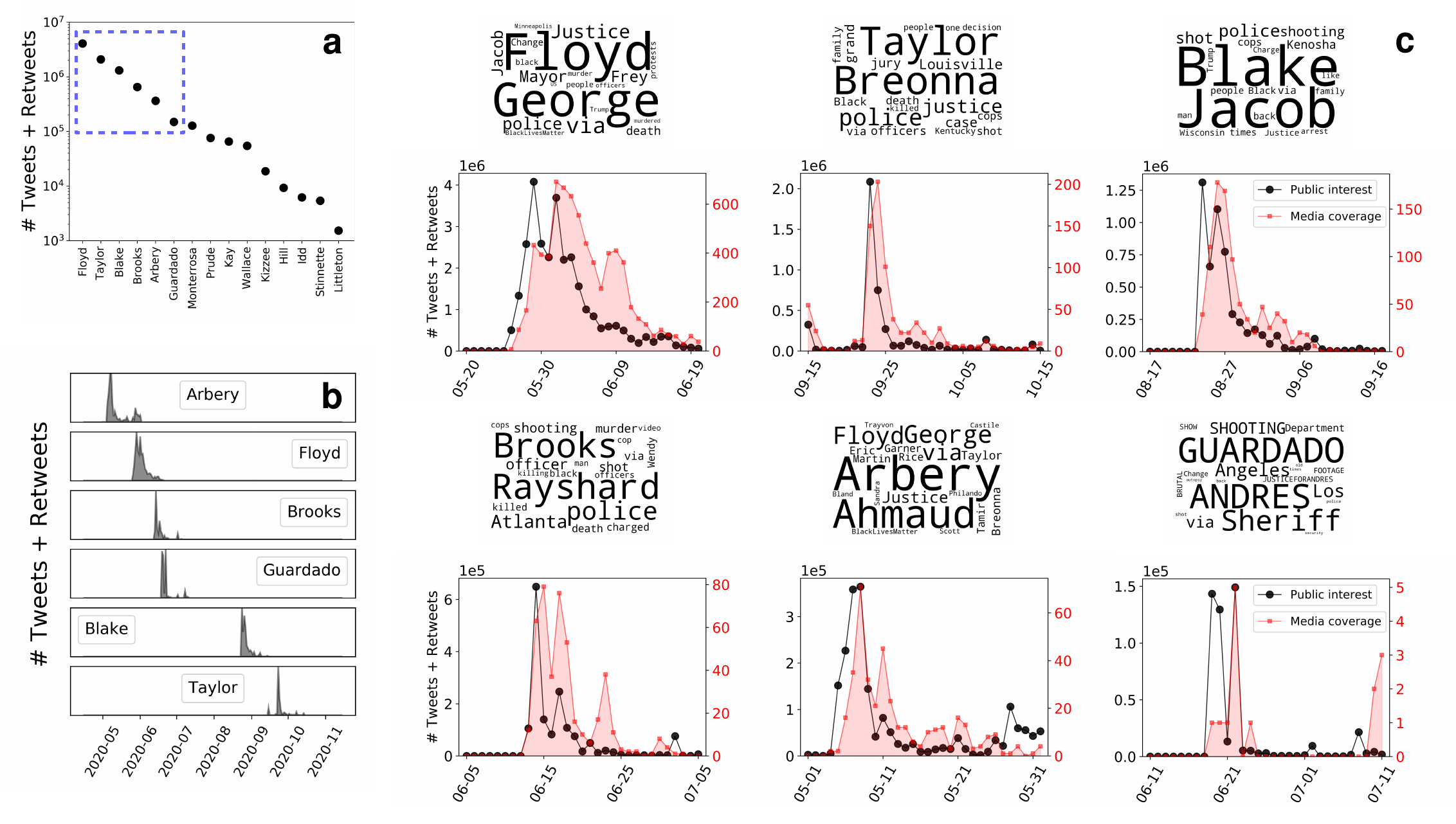}
    \caption{\textbf{Evolution of public interest and media coverage}. (a) Peak values of tweets and retweets for events related to the Black Lives Matter movement. Dashed square shows those with enough statistics, limited by the amount of media tweets (in the case of Guardado is about 10 tweets for the media accounts sampled). (b) Time traces of the volume of tweets and retweets, in chronological order. (c) Time traces of the volume of tweets and retweets (black circles, same data as panel (b)) and media accounts tweets (filled area). 
}
    \label{fig:topics}
\end{figure*}

\section*{Data}

The {\em Black Lives Matter} movement \cite{Carney2016} encompasses events of different nature and volume of activity in the social media (Figure \ref{fig:topics}a). Here we analyze a subset of the events well covered by media sources, as displayed in chronological order in Figure \ref{fig:topics}b. The time evolution of these events is shown in Figure \ref{fig:topics}c. Representing the public interest, we show in black the volume of tweets and retweets containing the keywords {\em George Floyd}, {\em Breonna Taylor}, {\em Jacob Blake}, {\em Rayshard Brooks}, {\em Ahmaud Arbery} and {\em Andr\'es Guardado}. Red filled curves correspond to the volume of tweets from the 29 most followed official media accounts containing the same keywords (See Supplementary Material for further details in the quantifying of media coverage). 

Besides a general resemblance of the public interest (black) and media coverage (red) across events, the traces are not merely copies of each other. One common feature is that the public interest grows faster than coverage at the events' onset. Here we propose that this effect is explained by the hightened social sensibility that characterizes these type of events. For this purpose we set up a model based on the one proposed by Granovetter, detailed in the next secion.

\section*{Model}

Our approach is grounded in the Granovetter model \cite{granovetter1978threshold}, originally proposed to explain the emergence of riots. In this model, agents adopt a binary state $s$ which we interpret as interested ($s=1$) or non-interested ($s=0$) in the event. The dynamics of the system is described in terms of the {\em public interest}, the fraction $p=\sum_i^N s_i/N$, where $N$ is the size of the system. Each agent is characterized by a threshold $\tau_i$, which is the fraction of interested agents needed to induce interest on the agent.
Thresholds are random variables whose cumulative distribution 
$S(p) = P(\tau<p)$ is interpreted here as {\em social engagement}, given that it represents the fraction of agents that become active due to their threshold lies below $p$. 

Assuming that thresholds are normally distributed $\tau\sim N(\mu,\sigma)$, we have:

\begin{equation}
     S(p|\mu,\sigma) =  \frac{1}{\sqrt{2\pi\sigma^2}}\int_{-\infty}^{p}e^{-\frac{(\tau-\mu)^2}{2\sigma^2}} d\tau.
         \label{eq:2}
\end{equation}
When $\mu$ is low, small groups can trigger the interest to the rest of the system. On the contrary, high values of $\mu$ would require a bigger fraction of interested people to induce interest to rest of the population. We therefore identify the quantity $1-\mu$ as the {\em social sensitivity} of the population.

In his original model, Granovetter described the dynamics of the public interest $p$ regardless of the influence of the media.
To include this, we propose a modified model that reads (see details in Materials and Methods):

\begin{equation}
    \frac{1}{\gamma}\frac{dp}{dt} = -p+e\,C(t)+(1-e)\,S(p|\mu(t),\sigma). 
    \label{eq:granovetter_mm}
\end{equation}

When the system is not exposed to the media ($e=0$), we recover the original Granovetter model, in which the dynamics of the public interest $p$ is driven by the social engagement $S$ with a time scale controlled by $\gamma$. On the contrary, when exposure to the media is maximum ($e=1$), the public interest is only driven by the media coverage $C$. In the general case $e\in (0,1)$, media coverage acts as an external field that modulates the public interest.
Of course, media coverage and public interest are far from being independent of each other. On the contrary, they feed one another; in mathematical terms, a closed model would require another equation for the evolution of media coverage modulated by the public interest. Here we tackle this by feeding equation \ref{eq:granovetter_mm} with the experimental time traces of media coverage $C(t)$.

Let us summarize the principal components of our model. On the one hand, we have two variables that quantify the volume of opinions and information shared by people: the public interest $p$ and the media coverage $C$. On the other hand, we have the social sensitivity $(1-\mu)$ and the social engagement $S$, two variables that describe macroscopic interactions among people. In the next section we show that the social variables can be reliably derived from the collected data shown in Figure 1c.  

\begin{figure*}[ht]
    \centering
    \includegraphics[width = \textwidth]{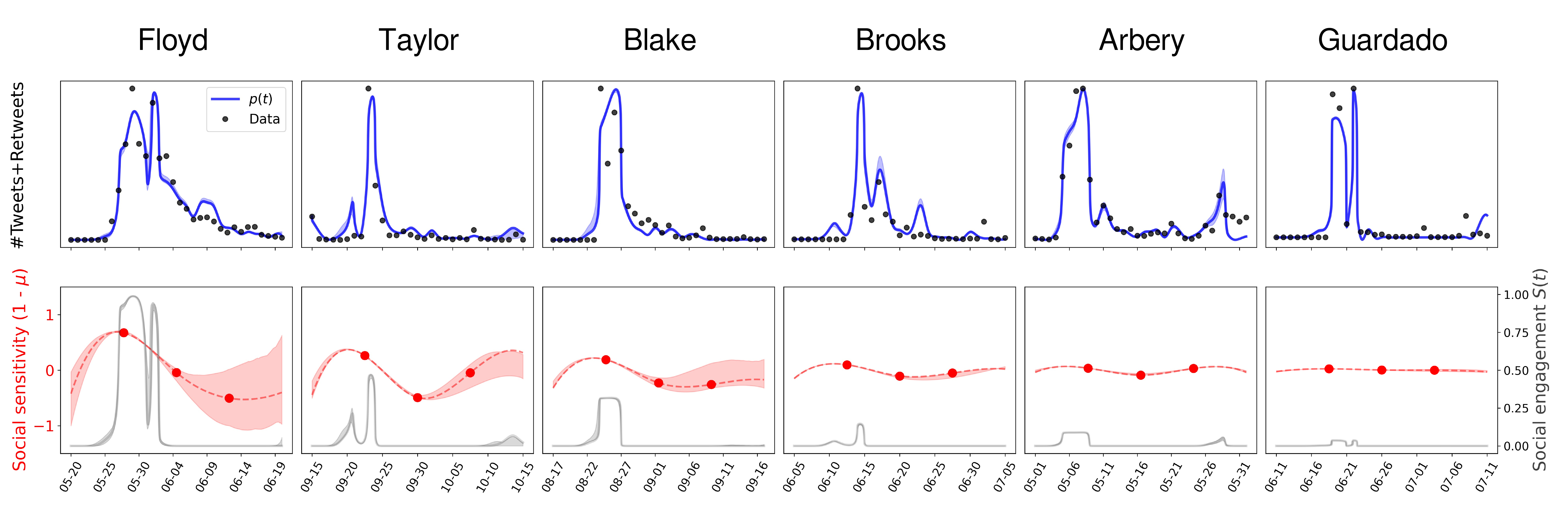}
    \caption{\textbf{Data fitting allows inferring social interactions.}
    Top panels. Points correspond to public interest  (tweets and retweets) along with the best fitting curves $p(t)$ (blue) obtained with the model of equations \ref{eq:granovetter_mm} and \ref{eq:2}. 
    Bottom panels. Social sensitivity $1-\mu(t)$, normalized to the event of mayor interest, which in this case is the murder of George Floyd. When the social sensitivity is high, more people become susceptible to the event. In grey lines the normalized social engagement $S(t) = S(p|\mu(t),\sigma)$ are also plotted.}
    \label{fig:model_integration}
\end{figure*}

\section*{Results}

To reconstruct the social variables, we integrate the equation \ref{eq:granovetter_mm} using the volume of twitted news as a proxy for the coverage $C(t)$. We seek for the functions $S(t)$ that minimize the difference between the resulting public interest and the volume of tweets and retweets. In Figure \ref{fig:model_integration} show the best fitting curves for the public interest (upper panels) and the reconstructed social engagement and social sensitivity (lower panels, grey and red curves respectively).

The two social variables are of a different nature. In fact, while the engagement $S(t)$ is a threshold-based variable whose dynamics can be expected to be fast, $1-\mu(t)$ represents the slower, more gradual build-up of social sensitivity across the whole population. Accordingly, we find that this variable changes appreciably over periods of $\sim$ 15 days which is, as expected, longer than the typical time scales of the media coverage and public interest (see Methods).

A summary of the fitting parameters is found in Table \ref{tab:table}. We find that exposure is rather stable across events, $e\sim0.4$. This says that, although media coverage is important, people is mainly influenced by the social environment in this kind of events. Different from exposure, the time scale $\gamma^{-1}$ decreases when the events accumulate over time. This is also expected, since the first four events (Arbery, Floyd, Brooks and Guardado) occured one immediately after the other (Figure \ref{fig:topics}b), speeding up the dynamics of public interest along the sequence. After a pause of about two months, the same speeding up effect is seen for Taylor, that occurred right after Blake.

\begin{table}[h]
\centering
    \begin{tabular}{p{1.5cm}|m{3cm}|m{2.5cm}}
         \bf  & \thead{Exposure $e$}  & \thead{Timescale $\gamma^{-1}$ \\ ($\times 10^{-3}$ days)}\\
         \hline
         \hline
         Arbery & $0.35$ $(0.31, 0.39)$ & $59$ $(36, 100)$ \\
         Floyd & $0.42$ $(0.40, 0.45)$ & $36$ $(22, 59)$ \\
         Brooks & $0.47$ $(0.47, 0.56)$ & $13$ $(12, 13)$ \\
         Guardado & $0.25$ $(0.25, 0.26)$ & $10$ $(10, 16)$\\
         Blake & $0.30$ $(0.26, 0.31)$ & $22$ $(17, 100)$\\
         Taylor & $0.47$ $(0.46, 0.54)$ & $13$ $(10, 16)$ \\
         \hline\hline
    \end{tabular}
    \caption{{\bf Fitted parameters}. Events are in chronological order (Figure \ref{fig:topics}). In all cases $\sigma = 0.2$. Intervals correspond to the 95\% confidence levels (see Data Fitting section for more details). }
    \label{tab:table}
\end{table}

\begin{figure}[ht]
    \centering
    \includegraphics[width = 7.5cm]{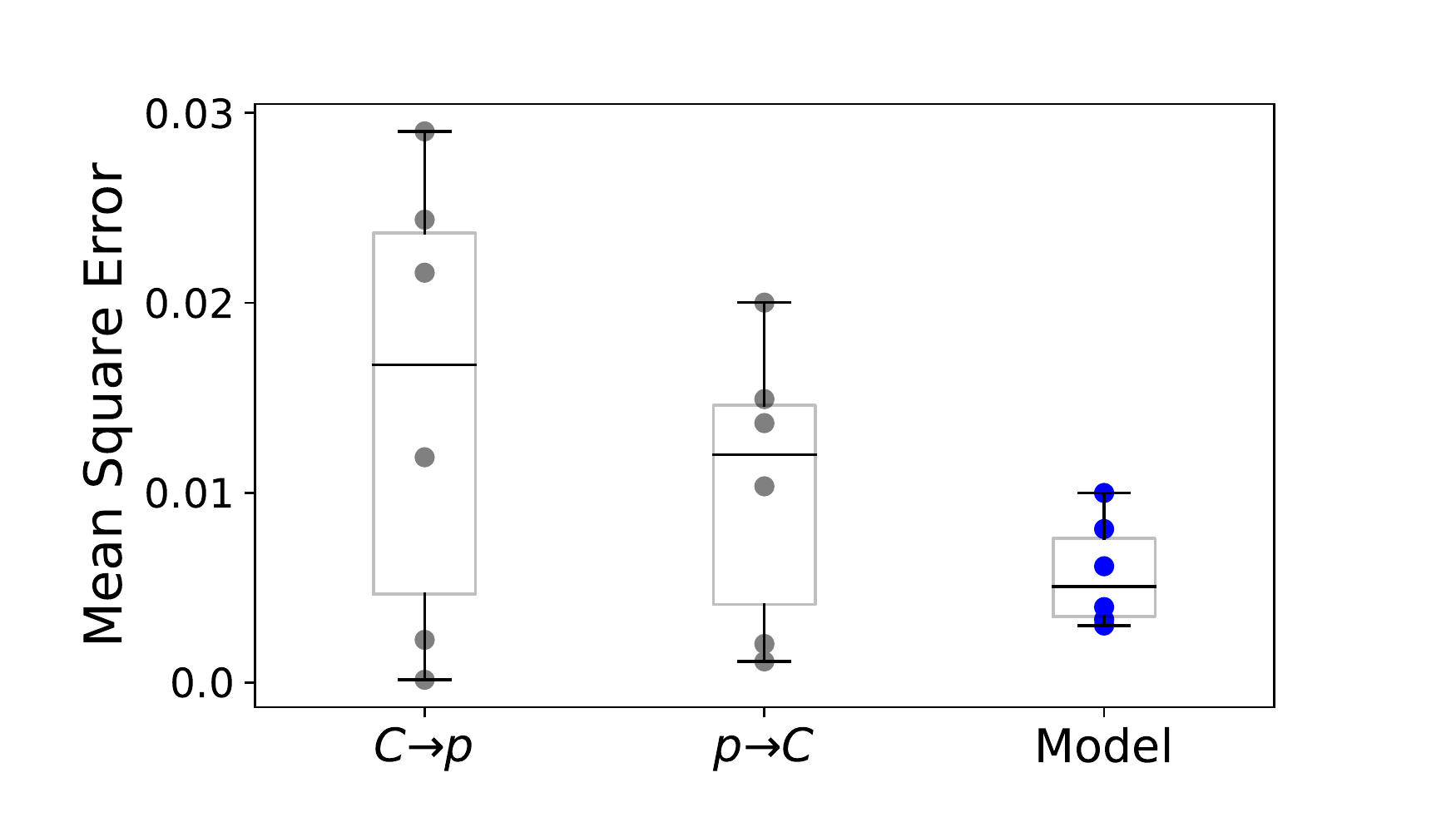}
    \caption{\textbf{Performace of the model.} We compare the goodness of fit with two basic models across the six events analyzed here. In one model, public interest alone predicts coverage ($p\to C$) and in the other, coverage alone predicts public interest ($C\to p$). Our model explains the data better than both basic models with the same number of fitting parameters.}
    \label{fig:null_model}
\end{figure}

To quantify the performance of our model, we compare its goodness of fit with two basic models: one in which coverage is predicted by public interest alone, and the opposite one where public interest is predicted by coverage alone (see Methods). In Figure \ref{fig:null_model} we show the mean square errors for the three models. Comparison of the basic models shows that public interest tends to predict coverage better than coverage predicts public interest. This is also apparent from the time series (Figure \ref{fig:topics}c), where the response of the media is slower with respect to the public interest at the onset of the events. Our model performs better than the basic models, explaining this delay by an increase in the social sensitivity $1-\mu(t)$. The inferred dynamics of the social variables are shown in the lower panels of Figure \ref{fig:model_integration}.  

\begin{figure*}[ht]
    \centering
    \includegraphics[width = \textwidth]{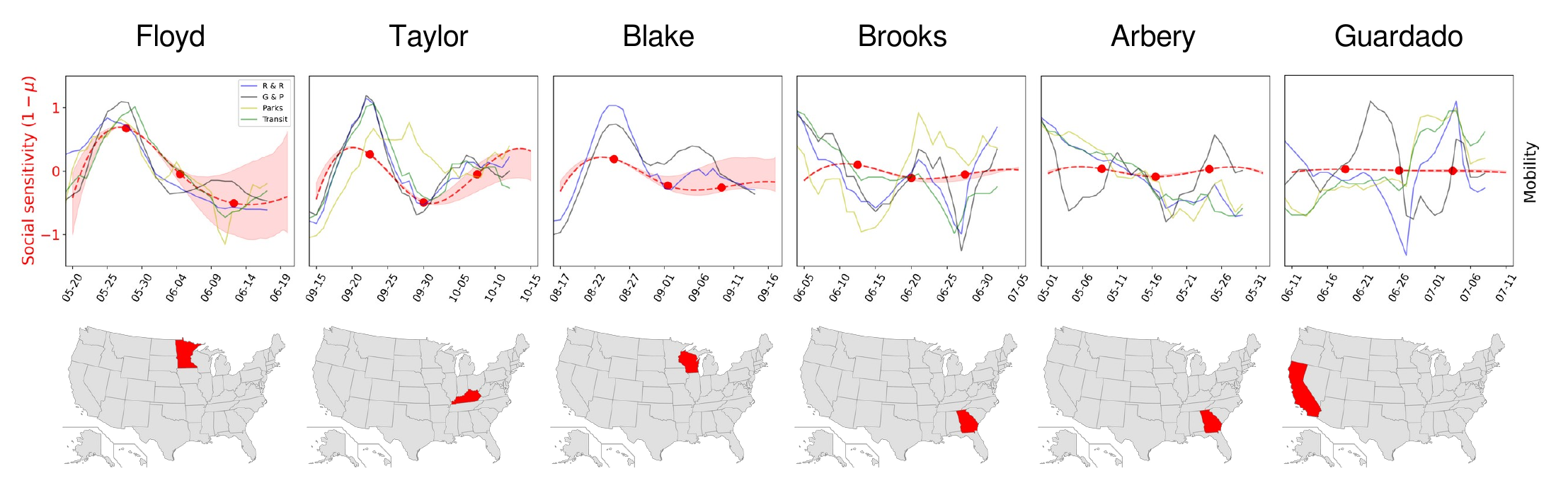}
    \caption{\textbf{Correlations between social sensitivity and mobility patterns.} Social sensitivity (red) and standardized mobility observables of the corresponding county. R \& R: retail and recreation; G \& P: groceries and pharmacies; Parks: public parks; Transit: transit in public transport stations (Parks and Transit not shown for Blake due to lack of data).
    All mobility measures were shifted $-3$ days and inverted for visualization purposes.}
    \label{fig:mobilization}
\end{figure*}

Bottom panels of Figure \ref{fig:model_integration} show periods of time of increasing social sensitivity, 
which leads to a sudden increase of the social engagement, when a macroscopic fraction of agents becomes interested in the events. If this dynamics is accurate, we should expect an impact beyond the digital environment.
To investigate the emergence of measurable collective activity associated to an increase in social sensitivity, we collected  mobility measures across the US territory \cite{google_mobility}. In Figure \ref{fig:mobilization} we show attendance to recreation places, groceries, pharmacies and public transport stations in the counties and periods of time when the events took place. We find different degrees of correlation between the social sensitivity  and mobility patterns for the most populous events using a lag of 3 days.
In the case of Floyd, social sensitivity correlates with all the four mobility measures, with a peak in the mean Spearman's rank coefficient $r = -0.82$; in the case of Taylor, $r=-0.47$ for two of the mobility measures; for Blake, $r=-0.48$ and only one measure ($p<0.05$ in all cases). The last three events were less massive, and we find no significant correlations with social sensitivity accordingly.

Taken together, these results suggest that our low-dimensional approximation of the Granovetter model captures the basic ingredients that regulate social responses of very different magnitudes, which are indeed capable of ignite social mobilizations. The model implements the hypothesis that agents become involved from media exposure and also from the presence of a critical mass of interested agents in the system, which leads to characterize the social sensitivity of the population. 

\section*{Discussion and conclusions}

Fluctuating interactions among people in massive social events are difficult to quantify. In this work we set up a simple mathematical model that allows us to infer how social interaction influence volume content representing public interest knowing media coverage. We then test our model on Twitter volume data related to the Black Lives Matter movement. 

We find that this formulation fits the experimental series better than two models in which public interest and coverage explain each other, in absence of social interactions. Crucially, we show that the evolution of the social sensibility correlates with variations in mobility data due to protests and riots during the events that draw the majority of the atention, presumably the most moving ones.

A possible limitation of our model is related to the assumption of uniform mixing \cite{Girvan2019} in pairwise interaction, given that public interest time series were collected from Twitter, which is indeed highly structured. The topology of social networks plays a key role when dealing with opinions of different sign, which give rise for instance to the emergence of echo chambers \cite{Cinelli2021,Cota2019}. In our work, however, we are dealing with the volume of keywords, regardless of ideological leanings. We show that, at least for the highly sensitive events analyzed here, the structure of the network can be disregarded, in line with similar models that assume uniform mixing and succesfully explain the dynamics of time series related  to different hashtags in Twitter \cite{towers2015mass, muhlmeyer2019event, lorenz2019accelerating}. Simple as it is, our model provides direct and interpretable measures of social engagement.

We are witnessing a rapid development of algorithms that are capable of organizing massive amounts of data based on statistical relationships. However, this growth has not been matched with a development of dynamical models capable of generalize our knowledge \cite{Brunton2016}. We hope that this work contributes to our understanding of public interest, showing the potential of a simple model to explain social reactions within and outside the digital environment.

\section*{Materials and Methods}
\subsection*{Corpus of data}
\label{sec:data}
We collected all the available tweets in English containing the keywords George Floyd, Breonna Taylor, Jacob Blake, Rayshard Brooks, Ahmaud Arbery, Andr\'es Guardado, Sean Monterrosa, Daniel Prude, Deon Kay, Walter Wallace Jr., Dijon Kizzee, Andre Hill, Dolal Idd, Marcellis Stinnette and Hakim Littleton, in a period of one month around a significant event related to each topic. 

Tweets were collected using the Twitter API v2 \cite{twitterv2}.
We also collected the tweets with the same keywords from the group of most followed news accounts in Twitter \cite{media_accounts}: @cnnbrk, @nytimes, @CNN, @BBCBreaking, @BBCWorld, @TheEconomist, @Reuters, @WSJ, @TIME, @ABC, @washingtonpost, @AP, @XHNews, @ndtv, @HuffPost, @BreakingNews, @guardian, @FinancialTimes, @SkyNews, @AJEnglish, @SkyNewsBreak, @Newsweek, @CNBC, @France24\_en, @guardiannews, @RT\_com, @Independent, @CBCNews, @Telegraph. 
Twitter data is available at https://shorturl.ae/AcUge.

Mobility measures correspond to the US County associated to each event.
From all mobility-related time series we extract the trend to compare with social engagement.

We provide here a brief context of the analyzed events. George Perry Floyd Jr. was murdered by a police officer in Minneapolis (Ramsey County), Minnesota, on May 25, 2020.
Breonna Taylor was fatally shot in Louisville (Jefferson County), Kentucky, on March 13, 2020. On September 23, several protests occur after charging decision announced in Taylor's death.
Jacob S. Blake was shot and seriously injured by a police officer in Kenosha County, Wisconsin, on August 23.
Rayshard Brooks was murdered on June 12, 2020 in Atlanta (Fulton County), Georgia.
Ahmaud Arbery was murdered on February 23, 2020 in Glynn County, Georgia. The case became resonant after the viralization of a video about the shooting that derive his death on May 7.
Andrés Guardado was killed by a Deputy Sheriff in Los Angeles County, California, on June 18, 2020.

\subsection*{Data fitting}
\label{sec:mu_parametrization}

We first normalized both public interest $p$ and media coverage $C$ respect to their peak values. To find a timescale for the dynamics of the social sensitivity, we parameterized $\mu(t)$ as a cubic-spline of $N$ equally spaced nodes within a 1-month period. The fitting error either falls abruptly at $N = 5$ (Floyd and Blake) or does not change significantly in the range $4\le N\le9$ (Taylor, Brooks, Arbery and Guardado). We therefore fixed the value $N=5$, for which $\mu$ changes appreciably on a timescale of $\sim15$ days.

The media coverage was interpolated in order to obtain a continuous signal. Interpolation and numerical integration of equations 1 and 2 were performed with the library \emph{scipy} \cite{2020SciPy-NMeth}. 
Parameter fitting was performed using a grid-search in parameter $\gamma \in [10^{-1} - 120]$ in combination with a minimization routine for a the rest of the parameters ($e \in [0, 1]$ and nodes of $\mu \in [-1, 2]$).
The routine consists on integrating the model and varying the parameters until a convergence critera is reached. We used Sequential Least Squares Programming for bounded problems in \emph{scipy} to minimize the mean square error between the output of the model and data.

Confidence intervals provided in table \ref{tab:table} and showed in Figures \ref{fig:model_integration} and \ref{fig:mobilization} correspond to fitting solutions with an error up to 10\% of the best solution in each case, except for Taylor and Brooks, where solutions with a fitting error up to 50\% of the best solution were reported. 

\subsection*{Basic models}
We compare the goodness of fit with two basic models. In one of them, coverage is predicted by public interest $p\to C$ and in the other it is the other way around, $C\to p$. Both basic models were set up to be nonlinear functions  approximated by order 7\textsuperscript{th} polynomials, $C(t)=\sum_{n=1}^7 a_n\,p^n(t)$ and  $p(t)=\sum_{n=1}^7 b_n\,C^n(t)$, without zeroth-order term ($a_0 = b_0 = 0$).
In this way, the basic models match the number of fitting parameters of the model ($e$, $\gamma$ and $\mu(t_i)$, with $1<i<5$).

\subsection*{Analytical formulation of the model}
\label{sec:formulacion_analitica}

Equation 1 is an analytical approximation of the threshold-based model proposed by Mark Granovetter \cite{granovetter1978threshold} with the addition of an external field.
In this model, agents adopt a binary state $s$ which we interpret as interest ($s = 1$) or non-interest ($s = 0$) in a given topic. The dynamics of the system is described in terms of the fraction of interested agents $p=\sum_i s_i/N$, where $N$ is the size of the system. The agents have also an associated threshold $\tau_i$, which is the fraction of interested agents needed to induce interest on agent $i$.
The thresholds are random variables between $0$ and $1$ taken from a probability density $f(\tau$). On the other hand, the external field is introduced through a parameter $C\in [0, 1]$ independent of the state of the system. 

With these ingredients, the dynamics of the system is as follows: 
the fraction of interested agents $p$ can change because a random agent $i$ interacts with the media with probability $e$ and become interested ($s_i= 1$) in a given topic with probability $C$ or disinterested ($s_i=0$) with probability $1-C$; otherwise, with probability $1-e$, the agent observes the system.
In this last case, if the fraction of interested agents is greater than the threshold of the agent ($p\geq\tau_i$), then it becomes interested ($s_i=1$); otherwise, it becomes disinterested ($s_i=0$). Agents' state are synchronously updated, independently from their initial state.

Following \cite{akhmetzhanov2013effects}, we derive the analytical expression for the dynamics of $p$ shown in equation 1. Let $q(p_k,t)$ be the probability that the fraction of interested agents at time $t$ is $p_k = k/N$. The master equation for $q(p_k, t)$ is: 
\begin{align*}
    \frac{dq(p_k,t)}{dt} &= Q(1|p_{k-1})q(p_{k-1},t) +  Q(0|p_{k+1})q(p_{k+1},t) \\
    &- Q(1|p_k) q(p_k,t) - Q(0|p_k) q(p_k,t)
\end{align*}
where $Q(1|p_{k})$ y $Q(0|p_{k})$ are the transition probabilities that a given agent become interested or disinterested given $p_k$. 
These probabilities are given by:
\begin{align*}
    Q(1|p_k) &= (1 - p_k) [(1-e) S(p_k) + e C] \\
    Q(0|p_k) &= p_k [(1-e) (1 - S(p_k)) + e (1-C)]
\end{align*}
where $S(p_k)$ is the threshold cumulative distribution function $S(p_k) = \int^{p_k} f(\tau) d\tau$, which by definition is the fraction of agents whose threshold is below $p_k$ ($S(p_k) \equiv P(\tau < p_k)$).

In the limit of infinite population ($N \to \infty$), $p_k \to p$, where $p$ is now the fraction of interested agents and a continue variable $\in [0, 1]$. 
In this limit, the following approximations are taken:
\begin{align*}
    p_{k \pm 1} &\to p \pm \eta \\
    q(p_{k \pm 1}, t) &\to q(p,t) \pm \frac{\partial q(p,t)}{\partial p}\eta \\
    S(p_{k \pm 1}) &\to S(p) \pm \frac{\partial S(p)}{\partial p}\eta
\end{align*}
with $\eta = 1/N$. Replacing the above expressions in the master equation and neglecting terms of $\eta^2$ order, we obtain: 
\begin{equation*}
    \frac{\partial q(p)}{\partial t} = -\frac{\partial}{\partial p}[(-p + S(p) - e S(p) + e C) q(p,t)] \eta
\end{equation*}
For a well-defined initial condition, $q(p,0) = \delta(p-p_0)$ ($\delta(x)$ is the Dirac's delta) and re-scaling time $t \to Nt$, the solution of the above equation (pages 53-54 of \cite{gardiner1985handbook}) is given by:
\begin{equation*}
    \frac{dp}{dt} = - p + (1-e) S(p) + e C
\end{equation*}
In particular, if the thresholds are normally distributed with mean $\mu$ and dispersion $\sigma$, $S(p) \equiv S(p|\mu, \sigma)$. 
Finally, by adding a constant $\gamma$ that allows to adjust the time-scale, equation 1 is obtained.

Equation 1 has equilibria given by $p_{eq} = (1-e) S(p_{eq}) + e C$
The stability of these points is given by the sign of:
\begin{equation*}
    \frac{d\dot{p}}{dp}|_{p_{eq}} = [- 1 + (1-e) \frac{d S(p)}{dp}]_{p_{eq}}
\end{equation*}
where can be observed that the parameter $C$ plays no role in setting the stability.

As reference, we summarize here all the variables and parameters of the model mentioned during the manuscript:

\begin{table}[h]
\centering
    \begin{tabular}{|m{1.2cm}|m{2.8cm}|}
    \hline
         $p(t)$ & Public interest \\ 
         $C(t)$ & Media coverage \\ 
         $S(p(t))$ & Social engagement \\ 
         $1 - \mu(t)$ & Social sensitivity\\ 
         $e$ & Media exposure\\
         $\gamma$ & Timescale \\ 
    \hline
    \end{tabular}
    \caption{{\bf Variables and parameters of the model.}}
    \label{tab:variables_reference}
\end{table}

\vspace{1cm}

\section*{Acknowledgements}

This research was partially funded by the Universidad de Buenos Aires (UBA), the Consejo Nacional de Investigaciones Científicas y Técnicas (CONICET) through grant PIP-11220200102083CO, and the Agencia Nacional de Promoción de la Investigación, el Desarrollo Tecnológico y la Innovación through grant PICT-2020-SERIEA-00966.

\bibliography{biblio}

\newpage
\input{supplmat}


\end{document}

%% file: supplmat.tex
\section*{Supplementary Material}

\subsection*{Coverage estimation}

\begin{figure*}[ht]
    \centering
    \includegraphics[width = 14cm]{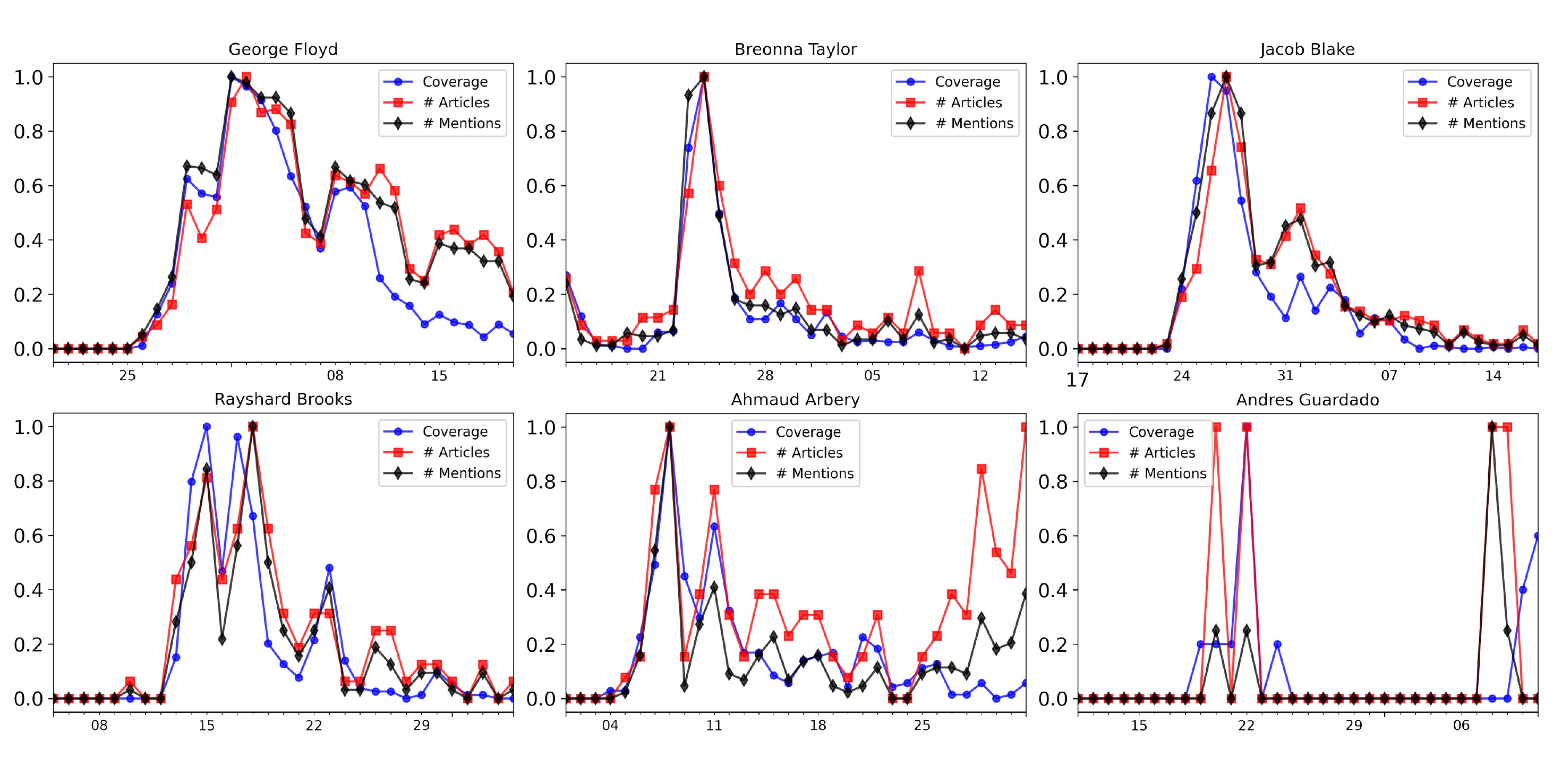}
    \caption{\textbf{Media coverage.} Coverage measured as twitter activity (coverage reported in the main manuscript, blue dots), number of articles (black diamonds) and number of mentions (red squares) from five main media outlets.}
    \label{fig:coverage}
\end{figure*}

Measures of media activity and public interest were collected from all the available tweets (in english language) containing the keywords George Floyd, Breonna Taylor, Jacob Blake, Rayshard Brooks, Ahmaud Arbery, Andr\'es Guardado, Sean Monterrosa, Daniel Prude, Deon Kay, Walter Wallace Jr., Dijon Kizzee, Andre Hill, Dolal Idd, Marcellis Stinnette and Hakim Littleton, in a period of one month around a significant event related to each topic using the Twitter API v2 \cite{twitterv2}.

\begin{figure*}[ht]
    \centering
    \includegraphics[width = 8cm]{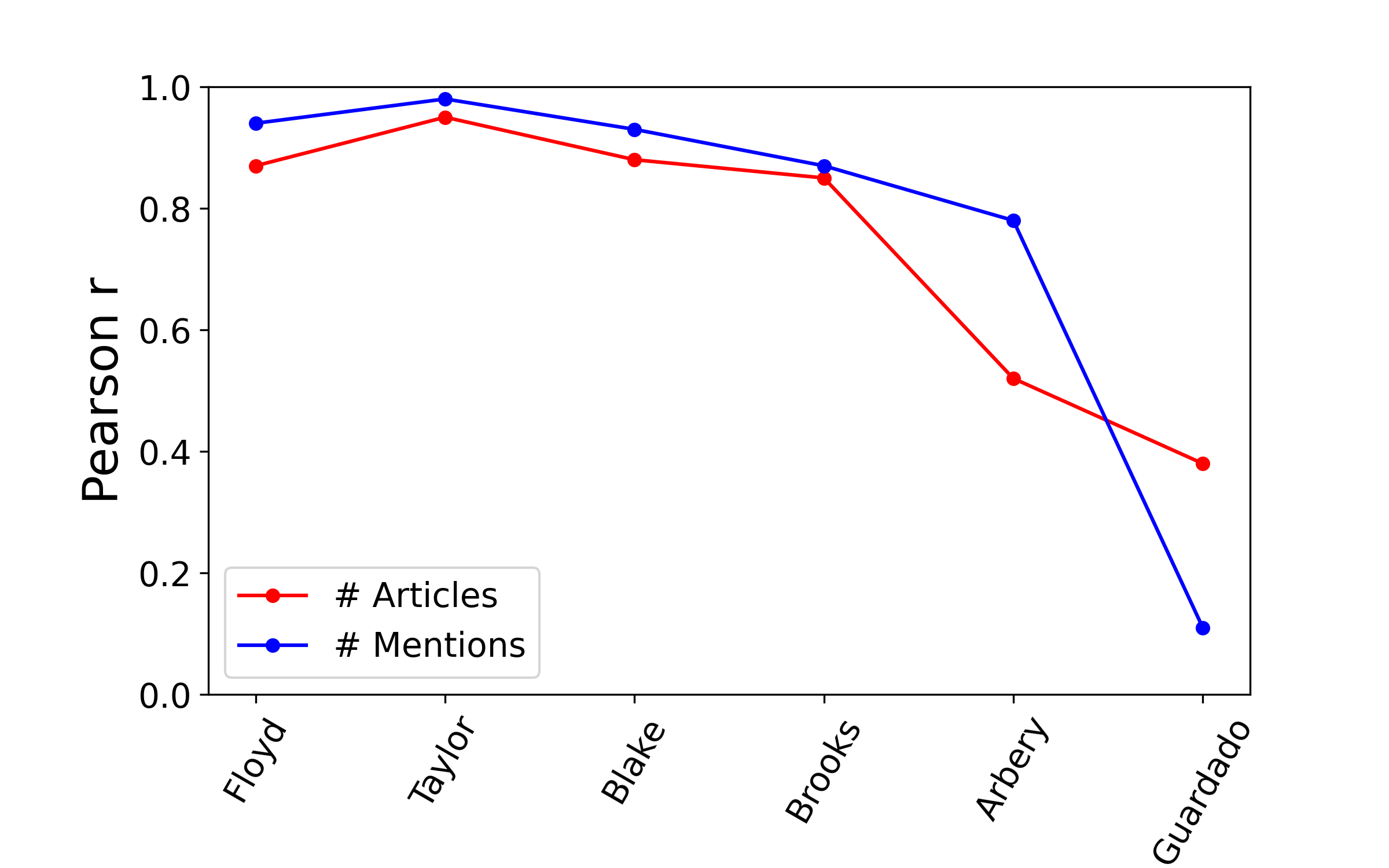}
    \caption{\textbf{Correlation measure.} Time evolution of Twitter activity is similar to the number of mentions in media articles.}
    \label{fig:correlation}
\end{figure*}

In particular, media coverage was estimated by collecting tweets with the mentioned keywords from the group of most followed news accounts in Twitter \cite{media_accounts}: @cnnbrk, @nytimes, @CNN, @BBCBreaking, @BBCWorld, @TheEconomist, @Reuters, @WSJ, @TIME, @ABC, @washingtonpost, @AP, @XHNews, @ndtv, @HuffPost, @BreakingNews, @guardian, @FinancialTimes, @SkyNews, @AJEnglish, @SkyNewsBreak, @Newsweek, @CNBC, @France24\_en, @guardiannews, @RT\_com, @Independent, @CBCNews, @Telegraph. 
Twitter data is available at https://shorturl.ae/AcUge.

To validate the measure of media coverage, we compare this quantity with information directly obtained from media articles.  In particular, we tracked news articles from five main media outlets such as The New York Times, Fox News, UsaToday, Washington Post and Huffington Post related to the main events analyzed in the paper. 

Figure \ref{fig:coverage} shows that the coverage reported in the main manuscript is  similar to the number of articles in which the keyword is mentioned and also with the number of mentions. 

Figure \ref{fig:correlation} shows the correlation between reported media coverage and the number of mentions in the articles. A coefficient higher than 0.8 is obtained in all cases, except from Guardado, suggesting that both approaches to measure media activity are equivalent.
The differences in the Guardado case is due to the fact that only a few articles were found in the analyzed media. 
